\documentclass[conference]{IEEEtran}

\usepackage{amsmath,amssymb,amsfonts,bm}
\usepackage{mathtools}
\usepackage{cite}
\usepackage{booktabs}
\usepackage{graphicx}
\usepackage{xcolor}

\usepackage[acronym]{glossaries}
\newacronym{scfde}{SC-FDE}{single-carrier frequency-domain equalization}
\newacronym{mmse}{MMSE}{minimum mean-square error}
\newacronym{mimo}{MIMO}{multiple-input multiple-output}
\newacronym{tdd}{TDD}{time-division duplex}
\newacronym{dmimo}{D-MIMO}{distributed MIMO}
\newacronym{ramimo}{RA-MIMO}{repeater-assisted massive MIMO}
\newacronym{isac}{ISAC}{integrated sensing and communications}
\newacronym{rsma}{RSMA}{rate-splitting multiple access}
\newacronym{sinr}{SINR}{signal-to-interference-plus-noise ratio}
\newacronym{noma}{NOMA}{non-orthogonal multiple access}
\newacronym{iot}{IoT}{Internet of Things}
\newacronym{ofdm}{OFDM}{orthogonal frequency-division multiplexing}
\newacronym{dft}{DFT}{discrete Fourier transform}
\newacronym{dftsofdm}{DFT-s-OFDM}{discrete Fourier transform-spread orthogonal frequency-division multiplexing}
\newacronym{ber}{BER}{bit-error rate}
\newacronym{snr}{SNR}{signal-to-noise ratio}
\newacronym{ue}{UE}{user equipment}
\newacronym{phy}{PHY}{physical layer}
\newcommand{\acp}[1]{\glspl{#1}}
\newcommand{\ac}[1]{\gls{#1}}

\newcommand{\E}{\mathbb{E}}
\newcommand{\Rea}{\operatorname{Re}}
\newcommand{\Ima}{\operatorname{Im}}
\newcommand{\diag}{\operatorname{diag}}

\newcommand{\vct}[1]{\boldsymbol{#1}}
\newcommand{\mat}[1]{\mathbf{#1}}

\AtBeginDocument{\setlength{\columnsep}{0.25in}}

\title{Repeater-Assisted MIMO Can Also Boost \\Frequency Diversity: A Semi-Analytic Study}

\author{\IEEEauthorblockN{Hiroki Iimori and Yuto Hama}\IEEEauthorblockA{Ericsson Research, Ericsson Japan K.K., Yokohama 220-0012, Japan\\
 (e-mail: \{hiroki.iimori, yuto.hama\}@ericsson.com)}}

\begin{document}
\maketitle

\begin{abstract}
Massive \ac{mimo} has enabled substantial spatial multiplexing and array gains in real-world systems, while \ac{dmimo} improves macro-diversity over wide areas at the cost of deployment complexity. \Gls{ramimo} is a lower-cost alternative that can recover key distributed-\ac{mimo} advantages. This paper asks whether repeater assistance can also enhance \emph{frequency diversity}. We study an uncoded \ac{dftsofdm} uplink with one-tap \ac{scfde} based on \ac{mmse} and derive a receiver-matched semi-analytic \ac{ber} expression by averaging over channel and interference realizations, without Gaussian approximation of residual despreading interference. The analysis clarifies how repeater delay reshapes frequency correlation, and waveform simulations confirm tight agreement with the derived expression together with improved high-\ac{snr} \ac{ber} decay, highlighting delay as a practical tuning knob.
\end{abstract}


\glsresetall

\section{Introduction}

Massive \ac{mimo} is a cornerstone technology of 5G and beyond cellular systems \cite{Marzetta2010,Larsson2014}. In sub-6~GHz \ac{tdd} deployments, reciprocity-based multi-user beamforming enables serving many \acp{ue} within the same time-frequency resource while achieving substantial array gain through phase-coherent combining, thereby improving signal quality and link budget. Despite these benefits, co-located massive \ac{mimo} may still suffer from coverage holes, and it can be challenging to reliably transmit multiple spatial streams to \acp{ue} when the propagation environment does not provide enough independent scatterers to yield sufficient channel rank. \Gls{dmimo}, also known as cell-free massive \ac{mimo}, has emerged as a promising remedy: by densely deploying access points over a wide area, \ac{dmimo} reduces path loss via shorter access-point-to-\ac{ue} distances, improves coverage, and mitigates shadow fading through enhanced macro diversity; moreover, favorable multi-path propagation can arise even under line-of-sight conditions \cite{Ngo2017,Ammar2022}.

While \ac{dmimo} offers compelling performance opportunities, its realization raises deployment and operational challenges. Key issues include transporting data to distributed units, deciding where processing should take place, synchronizing many access points, and distributing clocks and phase references with the accuracy needed for coherent multi-user downlink transmission; current standards and architectures are largely built around co-located \ac{mimo}, making a transition to originally-envisioned \ac{dmimo} costly. Motivated by the desire for smaller, backward-compatible steps toward densification, \ac{ramimo} deploys many small, inexpensive full-duplex repeaters that receive and retransmit signals instantaneously, thereby acting as \emph{active scatterers} (ordinary channel scatterers but with amplification) \cite{SaraComMag25}. Repeaters have long been used to improve coverage, and prior studies have examined capacity and coverage in repeater-aided \ac{mimo}, including outdoor-to-indoor improvements using full-duplex relays \cite{SaraComMag25}. In this paradigm, the base station retains control of multi-antenna processing while the repeaters enrich propagation, potentially narrowing the gap between co-located and distributed \ac{mimo} without backhaul or tight phase synchronization among many distributed transmitters.

Existing work on repeater-assisted \ac{mimo} spans a broad set of architectures and objectives, including repeater-assisted \ac{rsma}, \ac{isac} with repeater swarms, gain/phase optimization, dynamic-\ac{tdd} operation, full-duplex processing, activation control, and reciprocity calibration \cite{Vu2025RSMA,Jopanya2025SWARM,Iimori2023GLOBECOM,Andersson2026DynamicTDD,Mohammadi_FD_RA,Topal2025RA_MIMO,Chowdhury2026DualAntenna,Bai2026RepeaterSwarm,Le2025ARRNOMA,Larsson2024,Hara2026Calibration}. These studies report gains such as improved ergodic rate and fairness, better sensing performance under communication constraints, and behavior closer to distributed-\ac{mimo} without fronthaul. However, evaluations have focused mainly on spatial-domain metrics (coverage, spectral efficiency, \ac{sinr} uniformity, and interference management), while the effect of repeater assistance on frequency-domain channel structure and diversity remains comparatively underexplored. Since repeater-induced links add delayed and amplified components, they can reshape the effective impulse response and increase frequency selectivity, motivating the central question of this work: \emph{Can repeater assistance improve not only spatial macro diversity but also frequency diversity?}

To answer this question, this paper develops a receiver-matched analysis and validation framework for a wideband uplink with repeater-assisted propagation. Specifically, we consider an uncoded
single-carrier transmission implemented as \ac{ofdm} with DFT-precoding, known as \ac{dftsofdm}, with one-tap frequency-domain equalization based on \ac{mmse}
and derive a semi-analytic \ac{ber} expression by starting from the symbol-domain equivalent model and then taking the expectation over channel and interference realizations.
The main contributions are as follows: 1) we derive a semi-analytic \ac{ber} framework that preserves the exact structured residual interference term after despreading (without coarse Gaussian approximation); 2) we show analytically how repeater delays reshape frequency correlation through delay-domain PDP shifts; and 3) we validate the analysis against waveform-level Monte Carlo simulations and quantify the resulting high-\ac{snr} frequency-diversity gains. These results highlight repeater processing delay as a practical design parameter for reliability improvement.

\subsection{Notation}
Scalars are written in plain font (e.g., $x$, $h_k(\vct{\tau})$, $\gamma$), vectors in bold lowercase (e.g., $\vct{x}$), and matrices in bold uppercase (e.g., $\mat{F}$, $\mat{H}$). The operators $(\cdot)^T$, $(\cdot)^H$, and $(\cdot)^*$ denote transpose, Hermitian transpose, and complex conjugate. The real and imaginary parts are $\Rea\{\cdot\}$ and $\Ima\{\cdot\}$.
The symbol $*$ denotes linear convolution, and $\ell$ denotes the discrete-time tap (delay) index.

\section{System Model}
This section establishes the baseband system description used throughout the paper. To enable the frequency-diversity analysis and evaluation in later sections, we first present the transmit-receive signal model and its frequency-domain input-output relationship for a representative 
single-carrier implementation based on OFDM systems,
using \ac{dftsofdm} as a concrete example. We then introduce the channel model, including the direct path and the repeater-induced cascaded components.

\subsection{Input-Output Signal Model}

To keep the error-rate derivation and evaluation tractable while still exposing the essential frequency-diversity mechanism, 
we consider \ac{dftsofdm}, where
\ac{dft} precoding spreads each data symbol across the allocated subcarriers, and the subsequent IDFT (OFDM modulation) transforms the mapped spectrum back to the time domain. 
This structure produces a low-PAPR, single-carrier-equivalent
transmit waveform while retaining 
one-tap equalization in the frequency domain.

As a result of \ac{dft} spreading, each data symbol is distributed over frequency and therefore experiences multiple (generally different) channel responses, which enables capturing frequency diversity\footnote{Channel coding can further exploit this diversity gain, but it complicates the analysis; therefore, coded systems are left for future work.}.

Let $M\le N$ be the DFT spreading size and OFDM FFT size. Let $\vct{x}\in\mathbb{C}^{M\times 1}$ denote data symbols with $\E\{|x_i|^2\}=1$. The DFT-spread vector is
\begin{equation}
 \vct{x}_f=\mat{F}_M\vct{x},
 \label{eq:spread}
\end{equation}
where $\mat{F}_M$ is the unitary $M$-point \ac{dft} matrix.

Let $\mat{P}\in\{0,1\}^{N\times M}$ be the subcarrier-allocation matrix. The mapped frequency-domain \ac{ofdm} vector is $\vct{z}=\mat{P}\vct{x}_f$ and the $N$-sample time-domain transmit block is
\begin{equation}
 \vct{s}=\mat{F}_N^H\vct{z}
 =\mat{F}_N^H\mat{P}\mat{F}_M\vct{x}.
 \label{eq:time_domain_block}
\end{equation}

Next, CP insertion and removal are represented by matrices $\mat{C}_{\mathrm{cp}}$ and $\mat{R}_{\mathrm{cp}}$. With time-domain convolution matrix $\mat{H}_t(\vct{\tau})$,
\begin{equation}
 \vct{r}
 =\mat{R}_{\mathrm{cp}}\mat{H}_t(\vct{\tau})\mat{C}_{\mathrm{cp}}\vct{s}+\vct{w},
 \label{eq:cp_channel}
\end{equation}
where $\vct{w}$ is AWGN before FFT.

After FFT and subcarrier de-mapping,
the received symbol vector in the frequency-domain $\mathbf{y} \in \mathbb{C}^{M \times 1}$ is expressed as
\begin{equation}
 \vct{y}
 =\mat{P}^T\mat{F}_N\vct{r}
 =\mat{H}(\vct{\tau})\vct{x}_f+\vct{n},
 \label{eq:received}
\end{equation}
where $\mat{H}(\vct{\tau})=\diag(h_0(\vct{\tau}),\ldots,h_{M-1}(\vct{\tau}))$ and $\vct{n}\sim\mathcal{CN}(\vct{0},\sigma_n^2\mat{I}_M)$, assuming the cyclic prefix is longer than the effective channel impulse response so that inter-sample interference is avoided and the linear convolution is circular over the $N$-point block. Equation \eqref{eq:received} is the exact receiver-domain starting point used in the BER derivation.

\subsection{Repeater-Assisted Equivalent Channel}

This subsection describes how the effective channel is constructed when multiple repeaters assist the link. We analyze one equivalent data stream after spatial processing in a repeater-assisted massive-\ac{mimo} uplink, assuming perfect inter-layer/user interference cancellation\footnote{With large arrays, conventional \ac{mmse} spatial equalization can suppress interference to a large extent. Extensions to a full \ac{mimo} setup are left for future work.}. This stream is modeled as a DFT-s-OFDM link over an effective discrete-time channel. Let $R$ denote the number of repeaters. For repeater $r\in\{1,\ldots,R\}$, let
$h_{UR,r}[\ell]$ and $h_{RG,r}[\ell]$ denote the discrete-time impulse responses
of the UE$\to$repeater and repeater$\to$gNB channels, respectively. The
end-to-end cascaded impulse response associated with repeater $r$ is obtained by
convolving the two links,
\begin{equation}
 h_{c,r}[\ell]
 =\sum_{m} h_{UR,r}[m]\;h_{RG,r}[\ell-m]
 \triangleq (h_{UR,r} * h_{RG,r})[\ell],
 \label{eq:hc}
\end{equation}
which indicates that each repeater contributes one convolutional two-hop branch.

Let $h_d[\ell]$ denote the direct UE$\to$gNB impulse response, and let repeater
$r$ introduce a processing delay $\tau_r$ measured in integer samples. Defining
the delay vector $\vct{\tau}=[\tau_1,\ldots,\tau_R]^T$, the overall equivalent
impulse response seen at the receiver is the superposition of the direct and
delayed cascaded components,
\begin{equation}
 h[\ell;\vct{\tau}]
 =
 h_d[\ell]
 +
 \sum_{r=1}^{R} h_{c,r}[\ell-\tau_r].
 \label{eq:heff}
\end{equation}

This equivalent discrete-time channel is the impulse response experienced by
the DFT-s-OFDM waveform after propagation through all branches.
To connect the time-domain representation to the per-subcarrier model, the first
$N$ samples are stacked into the length-$N$ vector
\begin{equation}
 \vct{h}(\vct{\tau})
 =
 \big[
 h[0;\vct{\tau}],\,
 \ldots,\,
 h[N\!-\!1;\vct{\tau}]
 \big]^T,
 \label{eq:hstack}
\end{equation}
with implicit zero-padding beyond the channel support and its $N$-point DFT being then defined as
\begin{equation}
 \vct{h}_f(\vct{\tau})
 =
 \mat{F}_N \vct{h}(\vct{\tau})\in\mathbb{C}^{N\times 1}.
 \label{eq:hdft}
\end{equation}

Using the DFT time-shift property, a delay of $\tau_r$ samples in time produces
a multiplicative phase term $e^{-j2\pi k\tau_r/N}$ on the $k$th subcarrier.
Define the per-subcarrier composite channel coefficient as $h_k(\vct{\tau})\triangleq[\vct{h}_f(\vct{\tau})]_k$. Let $h_{d,k}$ and $h_{c,r,k}$ denote the
corresponding $N$-point DFT coefficients of $h_d[\ell]$ and $h_{c,r}[\ell]$.
Then
\begin{equation}
 h_k(\vct{\tau})
 =
 h_{d,k}
 +
 \sum_{r=1}^{R}
 e^{-j2\pi k\tau_r/N}\,h_{c,r,k},
 \quad k=0,\ldots,N-1.
 \label{eq:heff_freq}
\end{equation}

This expression shows that each repeater delay introduces a deterministic linear phase rotation across subcarriers, thereby modifying the coherent frequency-domain combining of the direct and repeater-assisted branches. Note that the channel is normalized such that adding repeaters does not increase channel power.

\section{Semi-Analytic BER Expression}
\subsection{Delay-Domain PDP and Subcarrier Correlation}
To provide intuition for why repeater delays can change frequency diversity, we briefly relate the composite impulse response in \eqref{eq:heff} to its frequency correlation. Define the branch-average power delay profiles (PDPs)
\begin{equation}
 p_d[\ell]=\E\{|h_d[\ell]|^2\},
 \qquad
 p_{c,r}[\ell]=\E\{|h_{c,r}[\ell]|^2\},
\end{equation}
and assuming zero-mean taps and statistical independence across the direct and cascaded branches, the equivalent average PDP
\begin{equation}
 p_h[\ell;\vct{\tau}]
 =
 p_d[\ell]
 +
 \sum_{r=1}^{R}
 p_{c,r}[\ell-\tau_r],
 \label{eq:pdp_general}
\end{equation}
which indicates that repeater delays shift the corresponding cascaded PDPs in the delay domain while preserving their average power.
To characterize frequency selectivity, define the frequency-domain channel correlation function as the subcarrier autocorrelation
\begin{equation}
 R_H[\Delta k;\vct{\tau}]
 \triangleq
 \E\!\left\{
 h_k(\vct{\tau})
 h_{k+\Delta k}^*(\vct{\tau})
 \right\}.
 \label{eq:freq_corr_def}
\end{equation}

Under the wide-sense stationary uncorrelated scattering (WSSUS) assumption, taps at different delays are uncorrelated and the correlation reduces to the DFT of the PDP,
\begin{equation}
 R_H[\Delta k;\vct{\tau}]
 =
 \sum_{\ell}
 p_h[\ell;\vct{\tau}]
 e^{-j2\pi \Delta k \ell/N}.
 \label{eq:rh_general}
\end{equation}

Equations \eqref{eq:pdp_general}-\eqref{eq:rh_general} highlight the key mechanism: by shifting energy in delay, $\vct{\tau}$ changes subcarrier correlation and therefore the amount of frequency diversity that can be extracted after equalization and despreading.

\subsection{MMSE Equalization and Despreading}
This subsection derives the symbol-domain detector output after applying 
one-tap
MMSE equalization in the frequency domain and then performing IDFT despreading. The result is a compact expression that separates the desired symbol term from the residual inter-symbol interference created by non-flat equalized gains across 
subcarriers
, plus the transformed noise. These expressions set up the exact decomposition needed to compute the conditional error probability in the next step, leading directly to \eqref{eq:symbol_split}--\eqref{eq:nu_vector} and then to \eqref{eq:gain_norm}.\par

The one-tap MMSE coefficient on the $k$th subcarrier is

\begin{equation}
 w_k=\frac{h_k(\vct{\tau})^*}{|h_k(\vct{\tau})|^2+\sigma_n^2},
 \label{eq:mmse_w}
\end{equation}
where noise amplification is suppressed by the denominator; at high SNR and strong channel gain, it approaches zero-forcing behavior; at low SNR or deep fades, it avoids excessive noise enhancement.

Collecting all coefficients in $\mat{W}=\diag(w_0,\ldots,w_{M-1})$, 
the resulting estimated symbol vector after DFT-despreading is

\begin{equation}
 \hat{\vct{x}}=\mat{F}_M^H\mat{W}\vct{y}
 =\mat{F}_M^H\mat{D}\vct{x}_f+\mat{F}_M^H\mat{W}\vct{n},
 \label{eq:xhat}
\end{equation}
where $\mat{D}=\diag(d_0,\ldots,d_{M-1})$ and $d_k=w_k h_k(\vct{\tau}) \in \mathbb{R}$
corresponding to the MMSE-equalized channel response.
This equation separates the detector output into an effective linear transformation of symbols and transformed noise.

\subsection{Gain Normalization and Symbol-Domain Decomposition}
This subsection converts the frequency-domain equalizer output into a symbol-domain model that is directly usable for \ac{ber} analysis. We first introduce a scalar gain normalization so that decisions are made at the original constellation scale, and then rewrite the combined MMSE+despreading operation as a circulant linear transformation $\mat{C}$ acting on the transmitted symbol vector. Exploiting the circulant structure yields an explicit decomposition of each detected symbol into the desired term, a structured self-interference term $I_i$, and an effective noise term, which is the key starting point for the conditional error-probability expressions derived next.\par

To align symbol decisions with the original constellation scale, 
we normalize by the average of the equalized gain
\begin{equation}
 g=\frac{1}{M}\sum_{k=0}^{M-1} d_k.
 \label{eq:gain_norm}
\end{equation}

Note that the MMSE-equalized channel response $d_k$ is always real-valued by definition of~\eqref{eq:mmse_w};
in numerical evaluations, a negligible imaginary part may remain due to finite-precision arithmetic.

Using \eqref{eq:gain_norm}, define
\begin{equation}
 \tilde{\vct{x}}=\frac{1}{g}\hat{\vct{x}}=\mat{C}\vct{x}+\vct{\nu},
 \quad
 \mat{C}=\frac{1}{g}\mat{F}_M^H\mat{D}\mat{F}_M,
 \label{eq:c_matrix}
\end{equation}
\begin{equation}
 \vct{\nu}=\frac{1}{g}\mat{F}_M^H\mat{W}\vct{n}.
 \label{eq:nu_vector}
\end{equation}

Since $\mat{D}$ is diagonal in frequency and surrounded by DFT/IDFT, $\mat{C}$ is circulant. Let $\vct{c}=[c_0,\ldots,c_{M-1}]^T$ be its first column,
\begin{equation}
 \vct{c}=\mathrm{IFFT}\!\left(\frac{d_0}{g},\ldots,\frac{d_{M-1}}{g}\right)^T.
 \label{eq:c_first_col}
\end{equation}

Then each detected symbol can be written as
\begin{equation}
 \tilde{x}_i=x_i+\underbrace{\sum_{m=1}^{M-1}c_mx_{i-m}}_{I_i}+\nu_i,
 \label{eq:symbol_split}
\end{equation}
where after equalization and despreading, each symbol equals the desired symbol plus structured finite-dimensional interference plus Gaussian noise.

\subsection{Post-Despread Noise Statistics and Modulation-General BER Form}
This subsection completes the probabilistic description of the post-despread detector output by (i) computing the variance of the transformed noise term $\nu_i$ after MMSE weighting, IDFT despreading, and gain normalization, and then (ii) expressing the symbol-to-symbol transition probability as an integral over the decision regions. With this, the BER can be written in a modulation-agnostic form conditioned on the structured interference term $I_i$ from \eqref{eq:symbol_split}. This provides the generic BER kernel that we will later average over channel and interference realizations, and it also sets up the QPSK specialization that follows.\par

From \eqref{eq:nu_vector},
\begin{equation}
 \sigma_\nu^2
 =\E\{\lvert\nu_i\rvert^2\}
 =\frac{\sigma_n^2}{g^2}\cdot\frac{1}{M}\sum_{k=0}^{M-1}|w_k|^2.
 \label{eq:nu_var}
\end{equation}

For a generic modulation alphabet $\mathcal{S}$ with $|\mathcal{S}|=M_c$ and $m_b=\log_2 M_c$ bits/symbol, define the bit-labeling map $\lambda:\mathcal{S}\rightarrow\{0,1\}^{m_b}$ and decision region $\mathcal{R}_{\hat{s}}$ for each $\hat{s}\in\mathcal{S}$. Conditioned on $I_i$ and transmitted symbol $s\in\mathcal{S}$, the transition probability is
\begin{equation}
 P(\hat{s}\mid s,I_i)
 =\iint_{z\in\mathcal{R}_{\hat{s}}}
 \frac{1}{\pi\sigma_\nu^2}
 \exp\!\left(-\frac{|z-(s+I_i)|^2}{\sigma_\nu^2}\right)\,dz.
 \label{eq:symbol_transition}
\end{equation}

The conditional BER for arbitrary modulation is then
\begin{equation}
 P_{b\mid I_i}^{(\mathcal{S})}
 =\frac{1}{m_b M_c}
 \sum_{s\in\mathcal{S}}
 \sum_{\hat{s}\in\mathcal{S}}
 d_H\!\left(\lambda(s),\lambda(\hat{s})\right)P(\hat{s}\mid s,I_i),
 \label{eq:pb_cond_general}
\end{equation}
where $d_H(\cdot,\cdot)$ is Hamming distance, and equations \eqref{eq:symbol_transition}--\eqref{eq:pb_cond_general} provide a modulation-agnostic semi-analytic BER core.

\subsection{QPSK Example}
This subsection instantiates the modulation-general BER framework of Step~3 for the practically important Gray-labeled QPSK case. By exploiting the separability of QPSK decisions into independent in-phase and quadrature real dimensions, the generic decision-region integral collapses to closed-form $Q(\cdot)$ expressions parameterized only by the structured interference realization $I_i$ and the post-despread noise variance. This produces an exact, low-complexity conditional BER expression that can be plugged directly into the outer expectation in the final semi-analytic averaging step.\par

For presentation clarity, we now specialize \eqref{eq:pb_cond_general} to Gray-QPSK. The same workflow applies to other constellations (e.g., $M$-QAM and $M$-PSK) by keeping \eqref{eq:symbol_transition}--\eqref{eq:pb_cond_general} and changing only decision regions and labeling.
For QPSK, each component takes values $\pm a$ with $a=1/\sqrt{2}$ and each real dimension has variance
\begin{equation}
 \sigma_r^2=\frac{\sigma_\nu^2}{2}.
 \label{eq:sigma_r}
\end{equation}

Conditioned on $I_i$, use the post-equalization model
\begin{equation}
\hat{x}_i = x_i + I_i + \nu_i,
\end{equation}
with Gray-QPSK symbols $x_i\in\{\pm a \pm j a\}$ and
\begin{equation}
\nu_i=n_I + j n_Q,\qquad n_I,n_Q\sim\mathcal{N}(0,\sigma_r^2),\qquad n_I\perp n_Q.
\end{equation}

Define
\begin{equation}
I_R \triangleq \Rea\{I_i\},\qquad I_Q \triangleq \Ima\{I_i\}.
\end{equation}

For the in-phase branch,
\begin{equation}
y_I=\Rea\{\hat{x}_i\}=x_I+I_R+n_I,\qquad x_I\in\{+a,-a\}.
\end{equation}
With zero-threshold detection:
\begin{align}
P(e_I\mid x_I=+a,I_i)
&=\Pr(y_I<0\mid x_I=+a,I_i) \nonumber\\
&=\Pr(n_I<-(a+I_R)) \nonumber\\
&=Q\!\left(\frac{a+I_R}{\sigma_r}\right),\\[1mm]
P(e_I\mid x_I=-a,I_i)
&=\Pr(y_I>0\mid x_I=-a,I_i) \nonumber\\
&=\Pr(n_I>a-I_R) \nonumber\\
&=Q\!\left(\frac{a-I_R}{\sigma_r}\right).
\end{align}
Since $\Pr(x_I=+a)=\Pr(x_I=-a)=\frac12$,
\begin{equation}
 P_{b,I|I_i}=\frac{1}{2}\left[
 Q\!\left(\frac{a+\Rea\{I_i\}}{\sigma_r}\right)
 +Q\!\left(\frac{a-\Rea\{I_i\}}{\sigma_r}\right)
 \right].
 \label{eq:pb_i_cond}
\end{equation}

Similarly, for the quadrature branch,
\begin{equation}
y_Q=\Ima\{\hat{x}_i\}=x_Q+I_Q+n_Q,\qquad x_Q\in\{+a,-a\},
\end{equation}
so
\begin{equation}
 P_{b,Q|I_i}=\frac{1}{2}\left[
 Q\!\left(\frac{a+\Ima\{I_i\}}{\sigma_r}\right)
 +Q\!\left(\frac{a-\Ima\{I_i\}}{\sigma_r}\right)
 \right].
 \label{eq:pb_q_cond}
\end{equation}

Therefore, conditioned on $I_i$,
\begin{equation}
 P_{b|I_i}=\frac{1}{2}\left(P_{b,I|I_i}+P_{b,Q|I_i}\right).
 \label{eq:pb_cond_qpsk}
\end{equation}

\subsection{Final Semi-Analytic \ac{ber} and Practical Estimator}

Since deriving the distribution of $I_i$ is generally challenging, and the expectation over $I_i$ is also difficult because the elements of $\vct{c}$ are correlated in practice, we resort to a semi-analytic \ac{ber} expression as follows.
The \ac{ber} at \ac{snr} $\gamma$ and repeater delay setting $\vct{\tau}$ is the expectation over channel randomness and interfering symbols:
\begin{equation}
 \bar{P}_b(\gamma,\vct{\tau})
 =\E_{\vct{h}(\vct{\tau})}\!\left[
 \E_{\{x_j\}_{j\neq i}}\!\left[P_{b\mid I_i}^{(\mathcal{S})}\mid\vct{h}(\vct{\tau})\right]
 \right].
 \label{eq:pb_final}
\end{equation}

If the QPSK specialization is used as exemplified in the previous subsection, $P_{b\mid I_i}^{(\mathcal{S})}$ is replaced by \eqref{eq:pb_cond_qpsk}. The term ``semi-analytic'' means the inner probability is analytical, while outer expectations are evaluated numerically.

To elaborate further, with $N_B$ batches, $N_H$ channel realizations per batch, and $N_S$ interference draws per channel,
\begin{equation}
 \hat{P}_b(\gamma,\vct{\tau})=\frac{1}{N_B}\sum_{b=1}^{N_B}
 \frac{1}{N_H}\sum_{\ell=1}^{N_H}
 \frac{1}{N_S}\sum_{q=1}^{N_S}P_{b\mid I_i^{(b,\ell,q)}}^{(\mathcal{S})},
 \label{eq:pb_estimator}
\end{equation}
which is the exact implementation counterpart of \eqref{eq:pb_final}; thus, analytical structure and simulation procedure are tightly aligned as shown in the simulation result section.

\section{Simulation Results}

\begin{table}[t!]
\caption{Simulation Setup}
\label{tab:sim_setup}
\centering
\begin{tabular}{ll}
\toprule
\textbf{Item} & \textbf{Setting} \\
\midrule
Waveform & DFT-s-OFDM \\
FFT size $N$ & $128$ \\
DFT size $M$ & $72$ \\
CP length & $32$ \\
Modulation & QPSK ($M_{\mathrm{QAM}}=4$) \\
Full-stack SNR grid & $0{:}1{:}25$ dB \\
Semi-analytic SNR grid & $0{:}1{:}45$ dB \\
Full-stack realizations per SNR & $100{,}000$ \\
Direct channel taps & $L_{\mathrm{d}}=4$ (equal PDP Rayleigh) \\
Repeater UE$\to$R taps & $L_{\mathrm{UR}}=6$ (equal PDP Rayleigh) \\
Repeater R$\to$gNB taps & $L_{\mathrm{RG}}=6$ (equal PDP Rayleigh) \\
Repeater-1 delay / gain & $d_1=8$, $g_1=1.0$ \\
Repeater-2 delay / gain & $d_2=14$, $g_2=1.0$ \\
Semi channel realizations per SNR & $3\times 10^7$ \\
Semi interference samples & $1000$ \\
Semi channel chunk size & $700$ \\
\bottomrule
\end{tabular}
\end{table}

This section evaluates \ac{ber} and diversity performance of \ac{dftsofdm} over 
i.i.d. Rayleigh fading with equal PDP
using both waveform-level Monte Carlo simulation (full-stack) and a semi-analytic \ac{ber} evaluation. Three channel configurations are considered: direct transmission only, direct plus one cascaded repeater branch, and direct plus two cascaded repeater branches with different processing delays. For each realization, the effective channel follows the superposition model in \eqref{eq:heff} and \eqref{eq:heff_freq}. The full-stack receiver applies frequency-domain MMSE equalization after CP removal and FFT, followed by IDFT despreading and QPSK demodulation. The semi-analytic BER is evaluated by averaging the conditional BER over channel and interference samples.
To characterize high-\ac{snr} reliability scaling, we compute the diversity metric directly from BER as
\begin{equation}
d(\gamma)\triangleq-\frac{\log P_e(\gamma)}{\log \gamma},
\end{equation}
where $\gamma$ denotes linear SNR and the diversity order is thoretically derived at $\gamma\rightarrow\infty$.

\begin{figure}[t!]
  \centering
  \includegraphics[width=\columnwidth]{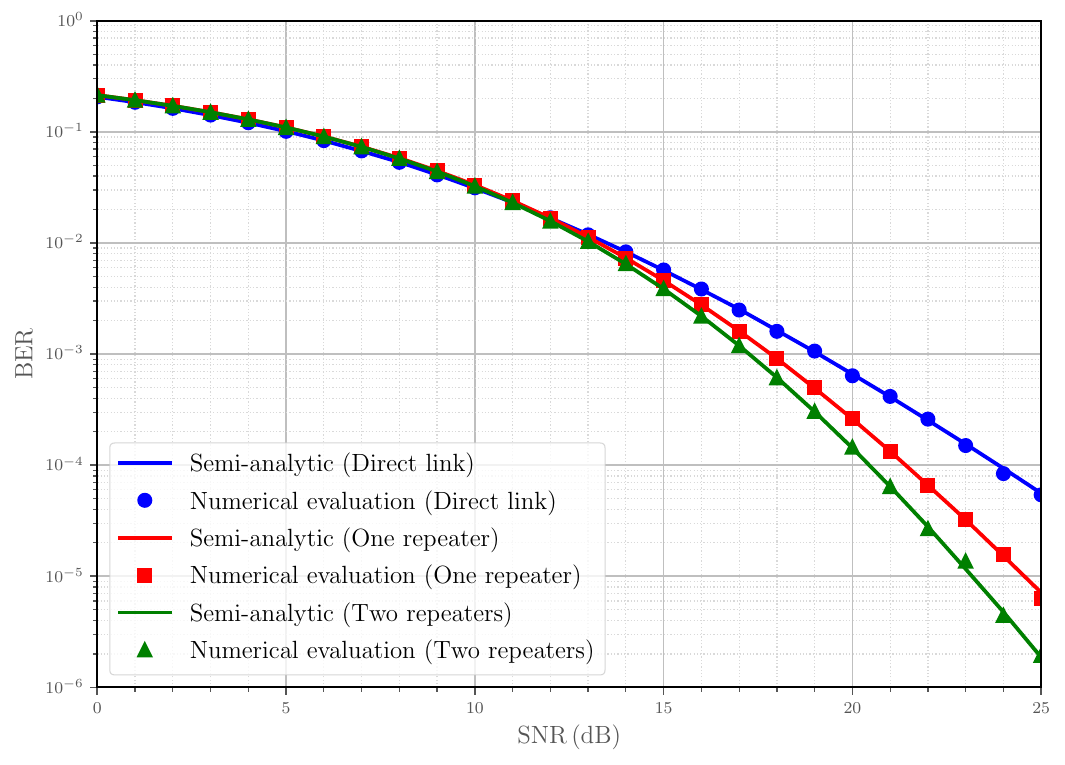}
  \caption{\ac{ber} comparison between the proposed semi-analytic evaluation and full-stack waveform Monte Carlo simulation.}
  \label{fig:ber_comparison}
\end{figure}

Table~\ref{tab:sim_setup} summarizes the simulation configuration. We use FFT size $N=128$, DFT spreading size $M=72$, and CP length $32$ with QPSK modulation. BER is computed on two SNR grids: $0{:}1{:}25$ dB for full-stack simulation and $0{:}1{:}45$ dB for semi-analytic evaluation. The direct channel uses $L_{\mathrm{d}}=4$ 
i.i.d. Rayleigh taps with equal PDP,
and each cascaded branch uses $L_{\mathrm{UR}}=6$ and $L_{\mathrm{RG}}=6$ taps with repeater settings $(d_1,g_1)=(8,1.0)$ and $(d_2,g_2)=(14,1.0)$. In the semi-analytic computation, $3\times 10^7$ channel realizations per SNR, $1000$ interference samples, and chunk size $700$ are used to balance estimator stability and runtime.

\subsection{Semi-Analytic and Waveform \ac{ber} Consistency}

Fig.~\ref{fig:ber_comparison} compares \ac{ber} versus \ac{snr} for the direct link, one-repeater link, and two-repeater link, where solid lines denote the semi-analytic results and markers denote full-stack Monte Carlo simulation. The near-overlap between each line-marker pair across the full \ac{snr} range validates the semi-analytic formulation as an accurate surrogate for waveform-level simulation. At low \ac{snr}, all three configurations are close because noise dominates performance; however, beyond roughly $15$ dB, repeater-assisted links show a clear advantage, with the two-repeater case consistently achieving the lowest \ac{ber} and the steepest high-\ac{snr} decay. For example, around $25$ dB the \ac{ber} is reduced from approximately $O(10^{-5})$ for the direct link to approximately $O(10^{-6})$ with one/two repeaters, and at a target \ac{ber} near $10^{-4}$ the repeater cases provide an \ac{snr} saving of roughly a few dB. These results indicate that adding cascaded repeater branches improves reliability and increases the effective diversity as shown below.

\subsection{Numerical Diversity Comparison}

\begin{figure}[t!]
  \centering
  \includegraphics[width=\columnwidth]{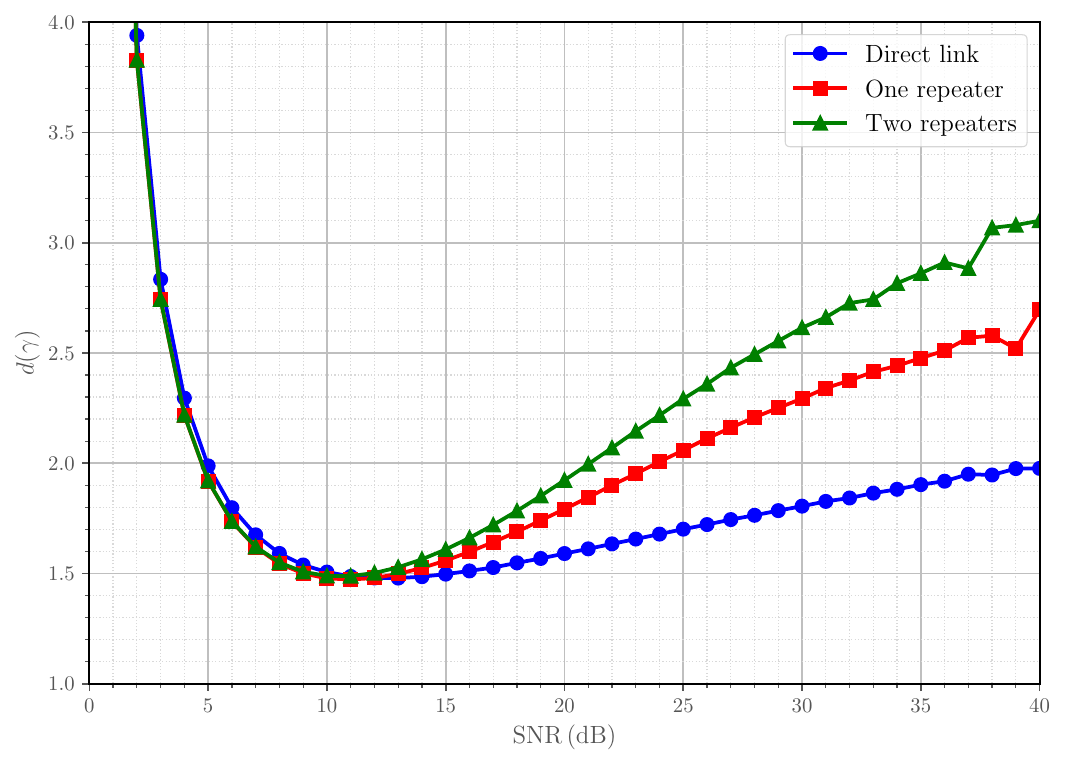}
  \caption{Numerical comparison of frequency-diversity behavior for the direct-only and repeater-assisted cases.}
  \label{fig:diversity_comparison}
\end{figure}

Fig.~\ref{fig:diversity_comparison} shows the numerically evaluated diversity metric $d_{\mathrm{dir}}(\gamma)=-\log(P_e)/\log(\gamma)$ for the direct, one-repeater, and two-repeater configurations. In this plot, $P_e$ is taken from the semi-analytic \ac{ber} curves (rather than full-stack Monte Carlo \ac{ber}), justified by the close agreement between semi-analytic and full-stack results established in Fig.~\ref{fig:ber_comparison}. All curves exhibit a similar low-\ac{snr} transient, which is not relevant when interpreting diversity because the diversity order is defined in the asymptotic regime $\gamma\to\infty$. The curves then settle near $d\!\approx\!1.5$ around the mid-\ac{snr} region (roughly $10$--$12$ dB), after which clear separation appears: the direct link increases most slowly, the one-repeater case grows faster, and the two-repeater case achieves the largest slope-equivalent diversity at high \ac{snr}. By $40$ dB, the diversity metric is approximately $2.0$ (direct), $2.7$ (one repeater), and $3.1$ (two repeaters), indicating that cascaded repeater branches provide a tangible diversity gain and progressively stronger error-decay behavior.

\section{Conclusion}
This paper investigated repeater-assisted \ac{dftsofdm} links over frequency-selective Rayleigh fading using both full-stack waveform simulation and a validated semi-analytic \ac{ber} framework. The results show that repeater assistance provides not only the well-known spatial/macroscopic coverage benefits but also measurable frequency-diversity gains through delayed cascaded components. Specifically, repeater-assisted configurations achieve lower \ac{ber} and higher effective diversity order than the direct link in the high-\ac{snr} region, with stronger gains when additional repeater branches are introduced.
From an engineering perspective, the gains are most visible once the link enters the medium/high-\ac{snr} region where frequency selectivity, rather than noise dominance, governs error decay. The results also suggest that repeater delay planning is not merely an implementation artifact: it directly affects subcarrier correlation and the diversity extracted after MMSE equalization and despreading.
An important direction for future work is to quantify how practical constraints affect these gains in realistic deployments. In particular, cyclic-prefix limitations, repeater group delay, repeater noise amplification, and inter-path correlation may reduce, saturate, or even negate the diversity gains observed under idealized assumptions. Therefore, system-level evaluation that jointly models \ac{phy} processing, synchronization, channel estimation, delay-budget constraints, and network geometry is needed to identify operating regimes where repeater-assisted frequency-diversity gains remain robust.

\bibliographystyle{IEEEtran}
\bibliography{ref}

\end{document}